\documentclass[11pt]{article}
\usepackage{lscape}

\makeatletter
\def\ps@top{\let\@mkboth\@gobbletwo
     \def\@oddhead{\rm\hfil\thepage\hfil}\def\@oddfoot{}
     \def\@evenhead{}\let\@evenfoot\@oddfoot}
\def\@bibsetup{\itemindent=-\leftmargin}
\def\@citesep{; }
\def\@cite#1#2{({#1\if@tempswa , #2\fi})}
\def\@biblabel#1{\hfill}
\def\thebibliography#1{\section*{References\markboth
 {REFERENCES}{REFERENCES}}\list
 {[\arabic{enumi}]}{\settowidth\labelwidth{[#1]}\leftmargin\labelwidth
 \advance\leftmargin\labelsep
 \usecounter{enumi}\@bibsetup}
 \def\newblock{\hskip .11em plus .33em minus -.07em}
 \sloppy
 \sfcode`\.=1000\relax}
\renewcommand{\section}{\@startsection {section}{1}{\z@}{-3.5ex plus -1ex minus 
    -.2ex}{2.3ex plus .2ex}{\centering\large\bf}}
\renewcommand{\subsection}{\@startsection{subsection}{2}{\z@}{-3.25ex plus
    -1ex minus -.2ex}{1.5ex plus .2ex}{\centering\bf}}
\makeatother
\begin{document}
\begin{center}{\bf
Placing the Deep Impact Mission into Context: \\
Two Decades of Observations of 9P/Tempel 1 from McDonald Observatory} \\ [20pt]
A. L. Cochran$^1$, E. S. Barker$^2$, M. D. Caballero$^3$, \\
and J. Gy\"orgey-Ries$^1$\\ [5pt]
1: McDonald Observatory, University of Texas, 1 University Station C1402,
Austin, TX 78712 \\
2: NASA Johnson Space Center, KX/Orbital Debris Program Office, 
2101 NASA Pkwy, Houston, TX 77058 \\
3: Georgia Institute of Technology, Center for Nonlinear Science and
School of Physics, Georgia Institute of Technology, Atlanta, GA 30332 \\ [5pt]
Accepted for Icarus \\ [1in]
\end{center}

\begin{center}{\bf Abstract}\end{center}
We report on low-spectral resolution observations of comet 9P/Tempel 1
from 1983, 1989, 1994 and 2005 using the 2.7m Harlan J. Smith telescope
of McDonald Observatory.  This comet was the target of NASA's
Deep Impact mission and our observations allowed us to 
characterize the comet prior to the impact.  We found that the
comet showed a decrease in gas production from 1983 to 2005, with
the the decrease being different factors for different species.
OH decreased by a factor 2.7, NH by 1.7, CN by 1.6, C$_{3}$ by
1.8, CH by 1.4 and C$_{2}$ by 1.3.  Despite the decrease in 
overall gas production and these slightly different decrease
factors, we find that the gas production rates of OH, NH, C$_{3}$, CH and
C$_{2}$ ratioed to that of CN were constant over all of the apparitions.
We saw no change in the production rate ratios after the impact.
We found that the peak gas production occurred about two months prior
to perihelion.   Comet Tempel 1 is a ``normal" comet.


\newpage
\section{Introduction}
On 4 July 2005 UT, comet 9P/Tempel 1 crashed into the Deep Impact spacecraft. 
The spacecraft was located in the path of the comet in order for the impact to
create a large crater, releasing material from the inside of the nucleus, to
enable study of interior ices.
In order to understand the changes which were
observed, it was necessary to have observations of the comet prior to
the impact.  In this paper, we detail low-resolution spectroscopic observations
obtained of comet 9P/Tempel 1 in its 1983, 1989, 1994 and 2005 apparitions 
using spectrographs on the 2.7m Harlan J. Smith telescope of McDonald
Observatory. 
The observations of 2005 include some made on the night of impact plus the two 
nights following the impact.

Comet 9P/Tempel 1 is a Jupiter family comet with a 5.5 year period.  Its
orbit is such that it has alternating good and poor viewing apparitions.
The apparitions 11 years apart have very similar viewing geometry.
Over the span of our observations, the perihelion date changed slightly
and the perihelion distance moved outwards from 1.49 to 1.51\,{\sc au}
(see Table~\ref{orbit}).  All of our data were obtained pre-perihelion
with the exception of the data from 5 and 6 July 2005.
\begin{table}[h!]
\caption{Orbital Parameters}\label{orbit}
\vspace*{10pt}
\centering
\begin{tabular}{ccccc}
\hline
Year & Perihelion & Perihelion & $e$ & $i$ \\
     & Date & Distance ({\sc {\sc au}}) & & (degrees) \\
\hline
1983 & Jul. 9.797 & 1.491117 & 0.520898 & 10.5571 \\
1989 & Jan. 4.443 & 1.496725 & 0.519684 & 10.5462 \\
1994 & Jul. 3.314 & 1.494151 & 0.520255 & 10.5518 \\
2000 & Jan. 2.617 & 1.500047 & 0.518953 & 10.5413 \\
2005 & Jul. 5.348 & 1.506434 & 0.517490 & 10.5303 \\
\hline
\multicolumn{5}{c}{Data from NASA/JPL's Horizons web interface} \\
\hline
\end{tabular}
\end{table}

\section{Observations}
The observations from 1983 were obtained with the Intensified Dissector 
Scanner (IDS) spectrograph.
This spectrograph utilized an intensification chain to 
rapidly build up the spectrum of two regions 52 arcsec apart and 4$\times4$
arcsec in size.  By moving the telescope under computer control, the user
could probe multiple positions in the coma.   The observations of 1983 have
already been published 
in Cochran {\it et al.} (1992). \nocite{cobarast92}  Table~\ref{idsdata} 
includes a log of the circumstances of these observations.
\begin{table}
\caption{The Intensified Dissector Spectrograph Observational Circumstances}\label{idsdata}
\vspace*{10pt}
\centering
\begin{tabular}{r@{ }l@{ }lccc}
\hline
\multicolumn{3}{c}{Date} & R$_h$ & $\Delta$ & Aperture \\
 & & & ({\sc {\sc au}}) & ({\sc {\sc au}}) & (km) \\
\hline
16 & Feb & 83$^*$ & 2.01 & 1.22 & 3532$\times$3532 \\
17 & Feb & 83$^*$ & 2.00 & 1.21 & 3496$\times$3496 \\
13 & Mar & 83$^*$ & 1.87 & 0.94 & 4073$\times$4073 \\
14 & Mar & 83$^*$ & 1.86 & 0.93 & 4056$\times$4056 \\
09 & Apr & 83$^*$ & 1.73 & 0.76 & 2199$\times$2199 \\
10 & May & 83     & 1.60 & 0.71 & 2071$\times$2071 \\
12 & May & 83     & 1.60 & 0.72 & 2077$\times$2077 \\
09 & Jun & 83$^*$ & 1.52 & 0.78 & 2269$\times$2269 \\
10 & Jun & 83     & 1.52 & 0.79 & 2280$\times$2280 \\
08 & Jul & 83     & 1.50 & 0.91 & 2640$\times$2640 \\
\hline
\multicolumn{6}{l}{* not photometric} \\
\hline
\end{tabular}
\end{table}

The observations of 1989, 1994 and 2005 were obtained with the Large 
Cassegrain Spectrograph (LCS), a long-slit CCD spectrograph.  The slit was
2 arcsec wide and 150 arcsec long, with the pixels each subtending 1.28\,arcsec
on the sky.  The slit could be rotated to any arbitrary angle
on the sky.  Observations typically covered 3000--5700\AA\ at 7\AA\ resolution.
Table~\ref{lcsdata} includes a log of the circumstances of these observations.
We often obtained more than one spectral image at any particular 
position angle of the slit.  In Table~\ref{lcsdata} the exposure
times are given for these multiple spectra (i.e. ``2$\times$1800\,sec" indicates
2 spectra of 30 min each or ``2700, 1800\,sec" indicates a 45 min
and 30 min spectrum).
\begin{table}
\caption{The Large Cassegrain Spectrograph Observational Circumstances}\label{lcsdata}
\vspace*{10pt}
\centering
\begin{tabular}{r@{ }l@{ }lccccccl}       
\hline
\multicolumn{3}{c}{Date} &  UT &
R$_h$  & $\Delta$ & \.{R} &  PA$_{sun}$ &
PA$_{slit}$ & exposure \\
 & & & Range & ({\sc {\sc au}}) & ({\sc {\sc au}}) & (km/sec) & (deg) & (deg) & (sec) \\
\hline
01& Jul & 89     & 09:43 -- 10:23 & 2.23  & 1.88  & 10.25 & 67.5  & 91     & 2400 \\
10& Feb & 94$^*$ & 11:06 -- 12:16 & 2.03  & 1.30  & -9.97 & 101.5 &  101.5 & 4200 \\
08& Mar & 94     & 09:03 -- 11:09 & 1.88  & 0.98  & -9.39 & 83.7  &  84    & 2$\times$3600 \\
07& Apr & 94     & 04:55 -- 06:49 & 1.73  & 0.75  & -8.13 & 14.5  & 104    &  2700, 3600 \\
12& Mar & 05     & 07:03 -- 08:06 & 1.87  & 0.95  & -9.26 & 77.5  &  90    & 2$\times$1800 \\
  &     &        & 08:20 -- 09:21 &       &       &       &       & 347.5  & 2$\times$1800 \\
  &     &        & 09:34 -- 11:11 &       &       &       &       & 77.5   & 3$\times$1800 \\
13& Mar & 05$^*$ & 10:01 -- 11:35 & 1.87  & 0.94  & -9.23 & 76.1  &  90    & 3$\times$1800 \\
06& Apr & 05     & 03:41 -- 04:45 & 1.75  & 0.77  & -8.23 & 13.0  &  90    & 2$\times$1800 \\
  &     &        & 04:55 -- 05:56 &       &       &       &       & 12     & 2$\times$1800 \\
  &     &        & 06:07 -- 07:09 &       &       &       &       & 57     & 2$\times$1800 \\
  &     &        & 07:19 -- 08:20 &       &       &       &       & 103    & 2$\times$1800 \\
07& Apr & 05$^*$ & 03:19 -- 04:21 & 1.74  & 0.77  & -8.18 & 9.7   &  90    & 2$\times$1800 \\
  &     &        & 05:53 -- 06:55 &       &       &       &       & 9      & 2$\times$1800 \\
  &     &        & 07:06 -- 08:08 &       &       &       &       & 99     & 2$\times$1800 \\
08& Apr & 05$^*$ & 05:03 -- 06:06 & 1.74  & 0.76  & -8.12 & 6.3   &  90    & 2$\times$1800 \\
  &     &        & 06:17 -- 07:20 &       &       &       &       & 6      & 2$\times$1800 \\
10& May & 05     & 03:30 -- 04:32 & 1.61  & 0.71  & -5.94 & 309.3 &  90    & 2$\times$1800 \\
  &     &        & 04:52 -- 05:55 &       &       &       &       & 130    & 2$\times$1800 \\
  &     &        & 06:05 -- 07:07 &       &       &       &       & 40     & 2$\times$1800 \\
12& May & 05     & 02:52 -- 03:43 & 1.60  & 0.72  & -5.77 & 307.8 &  90    & 1200, 1800 \\
08& Jun & 05     & 03:02 -- 03:52 & 1.53  & 0.78  & -3.15 & 296.1 &  90    & 900, 1200 \\
  &     &        & 03:51 -- 04:53 &       &       &       &       & 116    & 2$\times$1800 \\
  &     &        & 05:04 -- 06:04 &       &       &       &       & 26     & 2$\times$1800 \\
09& Jun & 05$^*$ & 03:16 -- 04:48 & 1.53  & 0.78  & -3.04 & 295.8 &  90    & 1800, 2$\times$700 \\
  &     &        & 04:53 -- 06:38 &       &       &       &       & 115    & 1500, 1800 \\
04& Jul & 05$^*$ & 06:07 -- 06:28 & 1.51  & 0.89  & -0.16 & 291.4 &  90    & 2$\times$600 \\
05& Jul & 05     & 03:34 -- 04:04 & 1.51  & 0.90  & -0.04 & 291.3 &  45    & 1800 \\
  &     &        & 04:27 -- 04:57 &       &       &       &       & 0      & 1800 \\
  &     &        & 05:09 -- 05:39 &       &       &       &       & 90     & 1800 \\
06& Jul & 05     & 03:27 -- 03:57 & 1.51  & 0.90  &  0.08 & 291.1 &  45    & 1800 \\
  &     &        & 04:20 -- 04:50 &       &       &       &       & 0      & 1800 \\
  &     &        & 04:57 -- 05:27 &       &       &       &       & 30     & 1800 \\
\hline
\multicolumn{10}{l}{* not photometric} \\
\multicolumn{10}{l}{PA = Position Angle, measured North through East} \\
\hline
\end{tabular}
\end{table}

Once the routine reductions were performed for each instrument (flat field, 
extraction, wavelength calibration and flux calibration for the IDS; bias
correction, flat field, 
wavelength and flux calibration for the LCS) we needed to compute the column 
densities and production rates for each spectrum.  For the IDS, this was
performed on the spectra from each individual position on the sky.  Since the
LCS is a long slit instrument, we 
would treat each row of the spectral image as a spectrum of a different 
position within the coma (or sometimes pairs of rows would be binned). 
First, we subtracted a sky spectrum from each comet spectrum, generally
weighting by the strength of the 5577\AA\ 
night sky O ($^1$S) line.  Next, we removed the solar spectrum which results
from the reflection of sunlight from the cometary dust. We used
either a solar analogue star spectrum obtained on the same night with
the same instrument or, when that was unavailable, a catalog solar spectrum
convolved with the instrumental slit profile.  A catalog
spectrum was always used for the IDS because the solar analogues are too bright.
The solar spectrum was typically weighted at five continuum wavelengths
in order to match the color of the cometary spectrum.   
Figure~1 of Cochran {\it et al}. (1992) demonstrates these steps.  For
observations of the OH band, we sometimes did not remove the solar or sky
spectrum because of the low signal/noise of these observations at the wavelength
of OH; removal introduced more noise into the comet spectrum than signal it
removed.

Once the sky and solar spectra were removed, we fit a continuum to the
region around each cometary emission band,
removed the continuum and then integrated
the flux above the continuum for each cometary molecular band observed.
(Though there should be no continuum left after the removal of the solar 
spectrum, in reality the color weighting of the solar spectrum is not perfect 
because there are few true continuum regions.  The fitted continuum
provides an accourate level to integrate above.)
The integrated fluxes were next converted to column densities using the
standard efficiency factors listed in Table~II of Cochran 
{\it et al} (1992).  In addition to the molecules listed in that table
(CN, C$_3$, CH, C$_2$ and NH$_2$), we used the ``Swings effect" calculations
of Schleicher and A'Hearn (1988) to derive a fluorescence efficiency for OH
and of Kim {\it et al} (1989) for NH.
\nocite{scah88,kiahco89}
At this point, for any particular night and slit orientation, we had multiple
values of column density as a function of position within the coma.  Finally,
we converted the data to production rates using the Haser (1957) model
\nocite{ha57} and the scale lengths listed in Table~\ref{scalelengths}. 
We have modified the outflow velocity law of Delsemme (1982) to
yield a velocity of 0.85\,km/sec at 1\,{\sc au} (as opposed to the original
0.58\,km/sec) to agree with the Giotto Halley results (see discussion
in Cochran and Schleicher 1993).  The velocity scales as $R_h^{-0.5}$
(R$_h$ = heliocentric distance).
\nocite{cosc93}
When there were multiple slit orientations in the course of a night, we
calculated separate production rates for each orientation.  
Since the slit always contained the optocenter of the comet, each orientation
should give the same value of the production rate if it was photometric. The
constancy of these values is an excellent indicator of our uncertainties.
\begin{table}[h!]
\caption{Adopted Haser Model Scale Lengths} \label{scalelengths}
\vspace*{10pt}
\centering
\begin{tabular}{lccc}
\hline
& Parent & Daughter & \\
& Scale length & Scale length & \\
Molecule & (km) & (km) & Reference$^a$ \\
\hline
OH & 2.4$\times10^4$ & 1.6$\times10^5$ & 1 \\
NH & 5.0$\times10^4$ & 1.5$\times10^5$ & 2 \\
CN & 1.7$\times10^4$ & 3.0$\times10^5$ & 3 \\
C$_{3}$ & 3.1$\times10^3$ & 1.5$\times10^5$ & 3 \\
CH & 7.8$\times10^4$ & 4.8$\times10^3$ & 4 \\
C$_{2}$$^b$ & 2.5$\times10^4$ & 1.2$\times10^5$ & 3\\
NH$_{2}$ & 4.1$\times10^3$ & 6.2$\times10^4$ &  \\
\hline
\multicolumn{4}{p{3.5in}}{$^a$References -- 1: Cochran and Schleicher (1993);
2: Randall {\it et al.} (1992); 3: Cochran (1986); 4: Cochran and Cochran (1990)}\\
\multicolumn{4}{l}{$^b$ C$_{2}$ parent scales as R$_{h}^{2.5}$, all others as R$_{h}^{2}$} \\
\hline
\end{tabular}
\nocite{co86haser}
\nocite{coco90,cosc93,raetal92}
\end{table}

Not all of the observations were obtained in photometric weather.  The weather
is noted in Tables~\ref{idsdata} and \ref{lcsdata}.  Since all wavelengths were
observed simultaneously, even non-photometric weather yields interesting
information because the relative abundances of the species are unaffected by
cloud (clouds have been shown to be grey (Wing 1967)).  \nocite{wi67}
Similarly, all positions
of a single spectral image of the LCS are obtained simultaneously, so the
spatial information is meaningful even when cloudy.  When there was
more than one spectral image at a position and the weather was
not photometric, the fainter (more cloudy) image was shifted upwards
to align with the brighter image.  In all cases except on 4 July
2005, individual spectral images were processed separately up
until the Haser model was fit; for 4 July, the spectra were averaged
first to increase the signal/noise (see discussion later about the use of
the Haser model for the July 2005 data).

Table~\ref{idsprodrates} lists the Haser model production rates
we derived from the 1983 data with the IDS.  These are fit to the
same data we reported in Cochran {\it et al.} (1992) except that
we have recalculated the Haser model production rates using the
higher outflow velocities described above.  We report production
rates for CN, C$_{3}$, C$_{2}$, CH and one data point for NH$_{2}$.
Note that the CN production rate for 9 April 1983 reported
in Cochran {\it et al.} (1992) was mis-typed.  The value in that
paper should have been log Q(CN)=24.96 instead of 24.21.  With
the higher velocity used in this work, we get log Q(CN)=25.13.

Tables~\ref{lcsprodrates1} and \ref{lcsprodrates2} give the production
rates for the data from 1989, 1994 and 2005.  We list
the production rates for each molecule observed (OH, NH, CN, C$_{3}$,
CH and C$_{2}$) on each night and at each slit orientation,  
as well as the binning in the
spatial direction (e.g. a bin of 2 means the effective
slit was 2.56 arcsec spatially
and 2 arcsec wide) and the number of positions which went into the
Haser model.   All of the underlying band intensities and 
column densities for these positions will be archived
in the Planetary Data System's Small Bodies Node (PDS SBN).

\begin{figure}
\vspace{4in}
\includegraphics{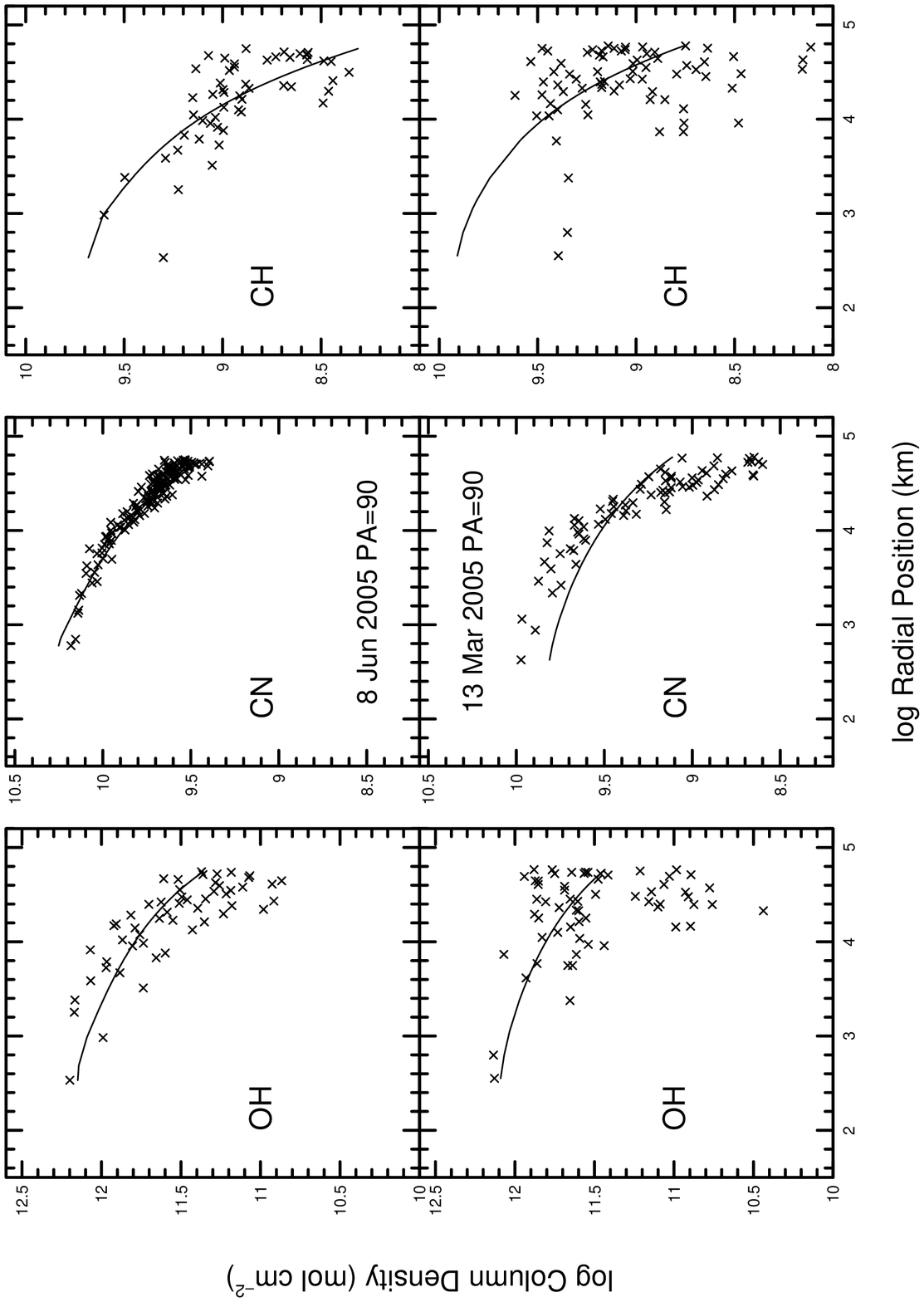}
\caption[fig1]{The column densities as a function of
cometocentric distance are shown for OH, CN and CH on 8 June 2005, when
it was clear, and 13 March 2005, when it was not.  The Haser model fits
are also plotted.  All boxes in a row are for the same date, marked
in the middle panel.  Clearly,
the data are better for some molecules than for others and are better
on 8 June than on 13 March.  Also, the Haser model does not always
fit.  See the text for a discussion of the errors.
}\label{columndensities}
\end{figure}

The data vary in quality as weather and cometary brightness changed
over an apparition.  Also, some molecules such as CN are intrinsically
easier to observe because they are strong lines; molecules such as CH are 
generally quite weak.  OH, while generally strong, is affected by low
quantum efficiency of the detector at 3080\AA\ and large atmospheric
extinction.  Figure~\ref{columndensities} shows examples of the 
column densities from individual coma positions for these three
molecules under excellent and poorer conditions.  The Haser
model fits are shown with the data.

Inspection of this figure shows several salient features.  Even under
the poorer conditions of 13 March 2005, the CN data show a clear trend.
However, the Haser model does not fit these data well.
Recall that
we are using fixed scale length values and not trying to fit the data.
This shows a weakness in the model but we use this approach for
inter-comparability of these data and those for other comets.  
We often see that the gas distribution for any molecule is dependent
on the orientation of the slit, with different profiles on either
side of the optocenter.  The Haser model does not take 
asymmetries into account.  

\begin{landscape}
\begin{table}
\caption{IDS Production Rates}\label{idsprodrates}
\vspace*{10pt}
\centering
{\small
\begin{tabular}{r@{ }l@{ }lrrrrrrrrrrrr}   
\hline
\multicolumn{3}{c}{Date} &
\multicolumn{1}{c}{log Q(CN)} & Npts &
\multicolumn{1}{c}{log Q(C$_{3}$)} & Npts &
\multicolumn{1}{c}{log Q(CH)} & Npts &
\multicolumn{1}{c}{log Q(C$_{2}$)} & Npts &
\multicolumn{1}{c}{log Q(C$_{2}$)} & Npts &
\multicolumn{1}{c}{log Q(NH$_{2}$)} & Npts \\
 & & & & & & & & & $\Delta v = 1$ & & $\Delta v = 0 $ \\
 & & & (sec$^{-1}$) &  & (sec$^{-1}$) & & (sec$^{-1}$) & & (sec$^{-1}$) & &
(sec$^{-1}$) & & (sec$^{-1}$) & \\
\hline
16 & Feb & 83 & 24.03 &  2  &       &   &        &    &        &      &       &   0 &       &     \\
17 & Feb & 83 & 24.48 &  2  &       &   &        &    &        &      &       &   0 &       &     \\
13 & Mar & 83 & 24.86 &  4  & 24.16 & 2 &        &    &        &      & 25.12 &   4 &       &     \\
14 & Mar & 83 & 24.53 & 18  & 23.45 & 6 &        &    &        &      & 24.86 &  14 &       &     \\
09 & Apr & 83 & 25.38 & 18  & 24.38 &16 &        &    & 25.26  &  15  & 25.10 &  15 &       &     \\
10 & May & 83 & 25.14 & 16  & 24.42 &14 & 25.23  &  6 & 25.06  &  16  & 25.11 &  16 & 24.75 &   2 \\
12 & May & 83 & 25.03 &  6  & 24.31 & 6 & 25.26  &  3 & 24.97  &   6  & 25.01 &   6 &       &     \\
09 & Jun & 83 & 24.70 &  4  & 23.77 & 2 &        &    &        &      & 24.82 &   4 &       &     \\
10 & Jun & 83 & 25.00 & 22  & 24.30 &14 &        &    & 25.09  &  15  & 25.01 &  19 &       &     \\
08 & Jul & 83 & 24.71 &  4  & 24.04 & 4 &        &    & 24.65  &   4  & 24.76 &   4 &       &     \\
\hline
\end{tabular}

}
\end{table}
\end{landscape}

\begin{landscape}
\begin{table}
\caption{LCS Production Rates (part 1)}\label{lcsprodrates1}
{\small
\vspace*{10pt}
\centering
\begin{tabular}{r@{ }l@{ }lrrlcrrlcrrlcrrlcr}      
\hline
\multicolumn{3}{c}{Date} & \multicolumn{1}{c}{PA} &
\multicolumn{2}{c}{log Q(OH)} & bin & Npts &
\multicolumn{2}{c}{log Q(NH)} & bin & Npts &
\multicolumn{2}{c}{log Q(CN)} & bin & Npts &
\multicolumn{2}{c}{log Q(CH)} & bin & Npts \\
 & & & (deg) & \multicolumn{2}{c}{(sec$^{-1}$)} & & & \multicolumn{2}{c}{(sec$^{-1}$)} & & &
\multicolumn{2}{c}{(sec$^{-1}$)} & & & \multicolumn{2}{c}{(sec$^{-1}$)} & & \\
\hline
  1& Jul & 89&      91 &  &       &   &      & &        &   &     & &  24.17 &   2&   13& &       &   &     \\
 10& Feb & 94&   101.5 &  & 26.44 &  2&    50& &  24.84:&  2&   46& &  24.38 &   1&  113& &  24.66:&  2&   34\\
  8& Mar & 94&      84 &  & 27.03 &  2&   121& &  25.15 &  2&  110& &  24.63 &   1&  243& &  25.31:&  2&  116\\
  7& Apr & 94&     104 &  & 27.23 &  1&   241& &  25.34:&  1&  241& &  24.88 &   1&  247& &  25.06 &  2&   97\\
 12& Mar & 05&      90 &  & 26.57:&  3&    48& &  24.81:&  3&   44& &  24.38 &   3&   62& &  24.86:&  3&   40\\
   &     &   &   347.5 &  & 26.51:&  3&    31& &  24.70:&  3&   42& &  24.33 &   3&   70& &  24.92:&  3&   33\\
   &     &   &    77.5 &  & 26.53:&  3&    41& &  24.83:&  3&   64& &  24.31 &   3&   93& &  24.75:&  3&   39\\
 13& Mar & 05&      90 &  & 26.86:&  2&   102& &  25.26:&  2&  104& &  24.43 &   2&  133& &  25.32:&  2&  119\\
  6& Apr & 05&      90 &  & 26.83 &  2&    79& &  25.10 &  2&  109& &  24.65 &   2&  118& &  24.83:&  2&   84\\
   &     &   &      12 &  & 26.97 &  2&   103& &  25.13 &  2&  116& &  24.67 &   2&  122& &  24.97:&  2&   87\\
   &     &   &      57 &  & 26.94 &  2&   106& &  25.11 &  2&  122& &  24.66 &   2&  112& &  24.82:&  2&   46\\
   &     &   &     103 &  & 26.93 &  2&   115& &  25.08 &  2&  120& &  24.63 &   2&  123& &  24.69:&  2&   55\\
  7& Apr & 05&      90 &  & 26.86:&  2&   102& &  25.15:&  2&  105& &  24.71 &   2&  125& &  25.09:&  2&  102\\
   &     &   &       9 &  & 27.02 &  2&   106& &  25.16 &  2&  110& &  24.69 &   2&  123& &  24.90:&  2&   81\\
   &     &   &      99 &  & 27.03 &  2&   111& &  25.17:&  2&  102& &  24.70 &   2&  124& &  24.97:&  2&   85\\
  8& Apr & 05&      90 &  & 26.97 &  2&   108& &        &   &     & &  24.64 &   2&  119& &  24.90:&  2&   79\\
   &     &   &       6 &  & 26.85 &  2&   106& &        &   &     & &  24.62 &   2&  121& &  25.02:&  2&   65\\
 10& May & 05&      90 &  & 27.07 &  2&   111& &  25.29 &  2&  112& &  24.90 &   1&  250& &  25.14 &  2&   89\\
   &     &   &     130 &  & 27.00 &  2&   113& &  25.23 &  2&  111& &  24.85 &   1&  248& &  25.07 &  2&   76\\
   &     &   &      40 &  & 26.91 &  2&    49& &  25.24:&  2&   84& &  24.89 &   1&  245& &  25.13 &  2&   75\\
 12& May & 05&      90 &  & 27.08 &  2&   117& &  25.26 &  2&  114& &  24.91 &   1&  248& &  25.12:&  2&   95\\
  8& Jun & 05&      90 &  & 26.78 &  1&    85& &  25.31 &  1&  121& &  24.75 &   1&  248& &  24.95:&  1&   84\\
   &     &   &     116 &  & 26.80 &  1&   122& &  25.25 &  1&  123& &  24.71 &   1&  246& &  24.76:&  1&   79\\
   &     &   &      26 &  & 26.62:&  1&    87& &  25.16 &  1&  117& &  24.73 &   1&  247& &  24.63:&  1&   52\\
  9& Jun & 05&      90 &  & 26.85:&  1&   119& &  25.29:&  1&  173& &  24.75 &   1&  371& &        &   &     \\
   &     &   &     115 &  & 26.78 &  1&    55& &  25.27 &  1&  111& &  24.59 &   1&  246& &        &   &     \\
  4& Jul & 05&      90 &  &       &   &      & &        &   &     & &  24.62 &   1&   43& &        &   &     \\
  5 &Jul & 05&      45 &  & 27.06:&  1&    76& &  25.27 &  1&  105& &  24.92 &   1&  118& &  25.09:&  1&   79\\
   &     &   &       0 &  & 27.14:&  1&    63& &  25.22 &  1&  101& &  24.86 &   1&  118& &  25.04:&  1&   82\\
   &     &   &      90 &  & 27.24:&  1&    38& &        &   &     & &  24.96 &   1&  113& &  25.03:&  1&   74\\
  6& Jul & 05&      45 &  & 27.02:&  2&    35& &  25.27 &  2&   53& &  24.92 &   1&  114& &  25.00:&  2&   48\\
   &     &   &       0 &  & 27.04:&  2&    32& &  25.19 &  2&   48& &  24.86 &   1&  114& &  24.99:&  2&   46\\
   &     &   &      30 &  & 26.86:&  2&    17& &  25.16:&  2&   47& &  24.82 &   1&  106& &  24.91:&  2&   43\\
\hline
\multicolumn{20}{l}{Note -- values with a : after them are very uncertain due to scatter; good to a factor of 2} \\
\hline
\end{tabular}

}
\end{table}
\end{landscape}

\begin{table}
\caption{LCS Production Rates (part 2)}\label{lcsprodrates2}
{\small
\vspace*{10pt}
\centering
\begin{tabular}{r@{ }l@{ }lrrlcrrlcrrlcr}      
\hline
\multicolumn{3}{c}{Date} & \multicolumn{1}{c}{PA} &
\multicolumn{2}{c}{log Q(C$_{3}$)} & bin & Npts &
\multicolumn{2}{c}{log Q(C$_{2}$)} & bin & Npts &
\multicolumn{2}{c}{log Q(C$_{2}$)} & bin & Npts \\
 & & & & & & & & \multicolumn{2}{c}{$\Delta v=1$} & & & \multicolumn{2}{c}{$\Delta v=0$} \\
 & & & (deg) & \multicolumn{2}{c}{(sec$^{-1}$)} & & & \multicolumn{2}{c}{(sec$^{-1}$)} & & &
\multicolumn{2}{c}{(sec$^{-1}$)} & & \\
\hline
  1& Jul & 89&      91 & &        &   &     &  &        &    &     &  &        &   &      \\
 10& Feb & 94&   101.5 & &  23.78 &  2&   47&  &  24.30:&  2 &   47&  &  24.54 &  1&   109\\
  8& Mar & 94&      84 & &  23.94 &  2&   79&  &  24.58 &  2 &  117&  &  24.80 &  1&   244\\
  7& Apr & 94&     104 & &  24.33 &  1&  232&  &  24.92 &  1 &  245&  &  24.98 &  1&   249\\
 12& Mar & 05&      90 & &  23.99 &  3&   48&  &  24.34:&  3 &   53&  &  24.18 &  3&    58\\
   &     &   &   347.5 & &  23.90 &  3&   54&  &  24.32:&  3 &   33&  &  24.13 &  3&    58\\
   &     &   &    77.5 & &  23.88:&  3&   63&  &  24.30:&  3 &   54&  &  24.04:&  3&    70\\
 13& Mar & 05&      90 & &  24.34 &  2&  131&  &  24.64 &  2 &   75&  &  24.50 &  2&   131\\
  6& Apr & 05&      90 & &  24.19 &  2&  107&  &  24.72 &  2 &  105&  &  24.75 &  2&   105\\
   &     &   &      12 & &  24.17 &  2&  107&  &  24.73 &  2 &  107&  &  24.77 &  2&   119\\
   &     &   &      57 & &  24.10 &  2&   99&  &  24.77 &  2 &  107&  &  24.74 &  2&   106\\
   &     &   &     103 & &  24.10 &  2&   98&  &  24.71 &  2 &  111&  &  24.75 &  2&   115\\
  7& Apr & 05&      90 & &  24.21 &  2&  109&  &  24.81 &  2 &   99&  &  24.81 &  2&   113\\
   &     &   &       9 & &  24.15 &  2&  104&  &  24.80 &  2 &  107&  &  24.81 &  2&   116\\
   &     &   &      99 & &  24.15 &  2&  104&  &  24.80 &  2 &  104&  &  24.82 &  2&   109\\
  8& Apr & 05&      90 & &  23.90 &  2&   41&  &        &    &     &  &  24.77 &  2&   112\\
   &     &   &       6 & &  23.84 &  2&   74&  &        &    &     &  &  24.73 &  2&   118\\
 10& May & 05&      90 & &  24.05 &  1&  114&  &  24.91 &  1 &  246&  &  25.00 &  1&   246\\
   &     &   &     130 & &  24.11 &  1&  169&  &  24.88 &  1 &  242&  &  24.98 &  1&   244\\
   &     &   &      40 & &  24.02 &  1&   72&  &  24.88 &  1 &  122&  &  25.00 &  1&   248\\
 12& May & 05&      90 & &  24.06 &  1&  140&  &  24.91 &  1 &  239&  &  25.03 &  1&   249\\
  8& Jun & 05&      90 & &  24.11 &  1&  194&  &  24.87 &  1 &  247&  &  24.85 &  1&   248\\
   &     &   &     116 & &  24.06 &  1&  202&  &  24.80 &  1 &  247&  &  24.83 &  1&   247\\
   &     &   &      26 & &  24.04 &  1&  198&  &  24.77 &  1 &  243&  &  24.82 &  1&   247\\
  9& Jun & 05&      90 & &  24.12 &  1&  303&  &  24.85 &  1 &  367&  &  24.84 &  1&   371\\
   &     &   &     115 & &  24.02 &  1&  174&  &  24.69 &  1 &  238&  &  24.77 &  1&   248\\
  4& Jul & 05&      90 & &  24.30:&  1&   44&  &  24.49 &  1 &   43&  &  24.71 &  1&    55\\
  5 &Jul & 05&      45 & &  24.14 &  1&  107&  &  24.96 &  1 &  117&  &  24.95 &  1&   118\\
   &     &   &       0 & &  24.11 &  1&  101&  &  24.91 &  1 &  116&  &  24.92 &  1&   115\\
   &     &   &      90 & &  24.21:&  1&   97&  &  24.87 &  1 &  114&  &  24.99 &  1&   118\\
  6& Jul & 05&      45 & &  24.13 &  1&  107&  &  24.95 &  1 &  117&  &  24.94 &  1&   115\\
   &     &   &       0 & &  24.11 &  1&  101&  &  24.92 &  1 &  115&  &  24.92 &  1&   115\\
   &     &   &      30 & &  24.07:&  1&   97&  &  24.88 &  1 &  113&  &  24.89 &  1&   118\\
\hline
\multicolumn{16}{l}{Note -- values with a : after them are very uncertain due to scatter; good to a factor of 2} \\
\hline
\end{tabular}

}
\end{table}

For OH and CH, even on a good night, there is much more scatter than
for CN.  Indeed, one can question how to interpret the ``fit" to the
data for CH on 13 March 2005!    We use a Monte Carlo approach to
quantifying our errors.  Each individual column density has 
an error based on the Poisson noise and on the quality of the
data (weather, airmass, strength of the feature).  We take the original data and
alter each individual column density by selecting from a normal distribution
with an initial width of the S/N at the peak and varying the width to account
for S/N decrease as we are farther from optocenter.  Then we rerun the model and
produce a new fit and production rate.  By repeatedly altering the data and
refitting them, we build up a picture of how errors in individual
data points affect the fit.  After 100 such runs, we find that
the data for CH on 13 March 2005 yields the same fit 
extremely reliably (log Q(CH) = 25.32$\pm$0.02).  This very
small error bar is not the result of the data being high quality (in
this case they are not) but is the result of us having 119
individual measures of the column density and, thus,  the errors in
individual points do not change the overall shape of
the gas distribution.  It is, of course, a
statistical error and does not include any systematic uncertainties
such as calibration problems or incorrect scale lengths or fluorescence
efficiencies. 
The innermost points for CH on this night are probably highly affected
by uncertainties in the removal of the continuum around this very
weak feature.  The dust continuum declines faster with cometocentric
distance than does the gas, so the effects of the continuum are
strongest near the optocenter.

In general, most of the CN data look closer in quality to the 8 June 2005
data than to the 13 March 2005 data (one of our poorest nights). 
The weaker features such as CH are
rarely of very high quality. Low quality, marginal data points are flagged
in the production rate tables.  The OH data quality depends on the airmass (so
is of lower quality near the end of the 2005 observations).  The overall
error bars are best judged by looking at the scatter in the column densities
and of the production rates derived for different slit orientations on
a single night.  Interested readers can obtain those from the PDS SBN. 
The scatter ranges from a few percent (e.g. 8 June 2005 CN) to large factors
(e.g. 13 March 2005 CH - but note that even with the scatter in CH, we
generally derived consistent production rates from different slit orientations).
For C$_{2}$, comparison of the results from the $\Delta v=1$ and
$\Delta v=0$ complexes give a good estimate of the accuracy. 
When multiple position angles of the slit were used, consistency of the
production rates is also a good indicator of accuracy.
The CN, C$_{3}$, and C$_{2}$ features are always the most certain values and
are probably good 0.1 dex.  Of course, on non-photometric nights, the
{\it absolute} values are not accurates but the {\it relative} values are.
The other systematic error is the fact that the Haser model cannot always fit
the profiles (NH is generally badly fit). 
This error is hard to quantify.  The Haser model, with
published scale lengths, was used as a standard tool for comparison.

\section{Trends in the Data}

\begin{figure}
\vspace{4in}
\includegraphics{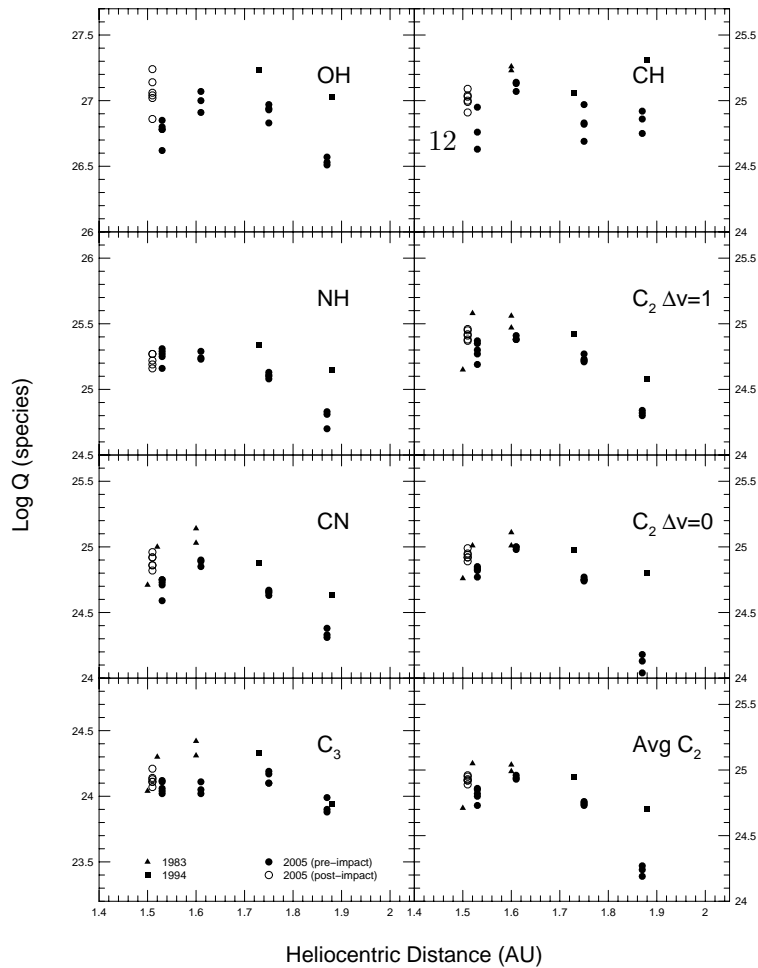}
\caption[fig2]{The production rates for nights with no
cloud are shown.  Data from 1983 are denoted with triangles, from 1994
with squares and from 2005 with circles.  The open circles are
the data from 5 and 6 July 2005, or post-impact.  Note that the
peak column densities are well before perihelion (excluding the
post-impact data).  This is true for all three apparitions.
The comet was systematically less productive in 2005.
}\label{prodplt}
\end{figure}

We have plotted the production rates of each molecule as a
function of heliocentric distance for nights on which it was
clear.  These are shown in Figure~\ref{prodplt}.  Several features
are apparent from this plot.  First, the systematic errors can be
estimated from the agreement between data points at the same heliocentric
distance in the 2005 data set.  As would be expected, the CN
values are much more consistent than the CH values.  All of the
data, except for the data from the two post-impact nights, were
obtained pre-perihelion.  However, the peak of the gas production
does not occur at the smallest heliocentric distances for any
molecule.  Instead, the production peaks prior to perihelion, when
the comet is at about 1.6\,{\sc au}, and
then declines (the CH and NH data are not of sufficient accuracy to judge
their behavior).  This behavior was first noted in data from 1983 by Osip
{\it et al.} (1992).
\nocite{osetal92}

There is a significant offset in the production of all of the
species from 1983/1994 to 2005 (unfortunately the weather
was not cooperative enough to determine if there is an offset
from 1983 to 1994).  For all molecules, the gas production
was higher in 1983/1994 than in 2005.  This was also reported
by Schleicher (2007).  \nocite{sc2007}
The magnitude of the decrease is different for different molecules.
We find that, at its peak, CN decreased by a factor of 1.6, C$_{3}$ by 1.8,
CH by 1.4, and C$_{2}$ by 1.3.  We were not able to observe NH
or OH in 1983 and did not observe the comet in clear weather
at its peak in 1994.  By extrapolating a linear trend for the 1994
data for clear nights, we can estimate a decrease of OH by a factor of 2.7
and of NH by a factor of 1.7.  The numbers are slightly
different than those factors seen by Schleicher (2007), who
saw decreases of OH by a factor of 2.5, NH and CN by 1.9, C$_{3}$ by 1.5
and C$_{2}$ by 1.4.  Our C$_{3}$ data at 1.6\,{\sc au} looks a little
low, so may explain the difference in our and Schleicher's C$_{3}$ decrease
factor.  We have no explanation why the C$_{3}$ is low except that the
model does not fit the gas distribution as well as at other times and
thus may be giving a false low value.

The data are somewhat sparse, with clumps of data at each of
several heliocentric distances.  The data were obtained approximately
monthly, based on the lunar cycle, and the viewing geometry 
was similar in the 1983, 1994 and 2005 apparition.  There is
a suggestion in Fig.~\ref{prodplt} that the behavior of the gas production of
some species with heliocentric distance was not the same in all 
three apparitions.  In particular, C$_{2}$, NH and perhaps CN seemed
to have higher production rates at 1.75\,{\sc au} relative to 1.9\,{\sc au}
in the 2005 apparition than in 1994.  However, the data were not
of the same quality in 1983 or 1994 as 2005 so this suggestion is
only tentative.

\begin{figure}
\vspace{4in}
\includegraphics{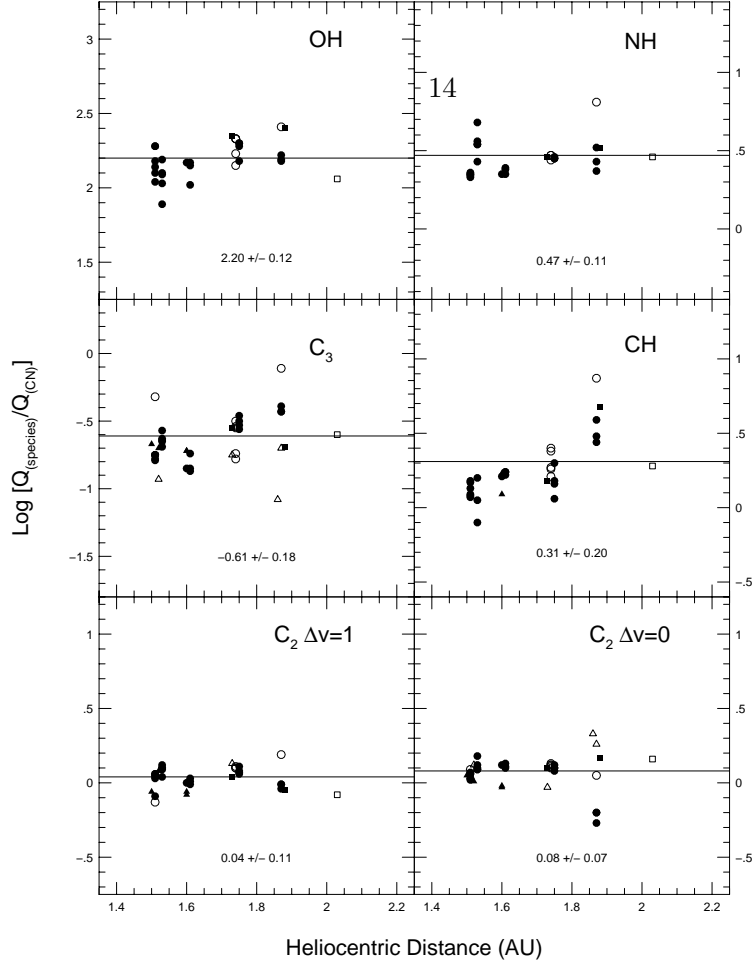}
\caption[fig3]{The production rates of various species
are ratioed to the production rate of CN and shown versus
heliocentric distance.  These ratios do not change with heliocentric
distance, with the possible exception of CH.
The average values are noted in each panel.   Data from 1983 are
triangles, from 1994 are squares and from 2005 are circles.  Filled
symbols are clear nights; open symbols are cloudy nights.  The circles
at 1.5\,{\sc {\sc au}} are the nights of 4--6 July 2005 UT, or the night
of impact and onwards.  The comet looks the same post-impact as before.
}\label{ratios}
\end{figure}

We can utilize all of the data, including the nights with clouds, to
look at gas production ratios.  These ratios have been used
by many authors to characterize comets as ``normal" or
``depleted" (A'Hearn {\it et al.} 1995; Cochran {\it et al.} 1992; 
Newburn and Spinrad 1984; Fink and Hicks 1996).
\nocite{ahetal95,cobarast92,nesp84,fihi96}
Figure~\ref{ratios} shows the production rates for OH, NH, C$_{3}$, CH,
and C$_{2}$ ratioed to that for CN, including all of the data.  Inspection
of this plot shows that these ratios were constant with
heliocentric distance with the possible exception of CH/CN.  However,
the CH data are not as high quality as other data since
this feature is extremely weak.  For the other species, while there
are obvious outliers, there appears to be no change in the 
production rate ratios with heliocentric distance.  The mean
values are denoted in each panel of the plot.  Tempel 1 looks
to be a normal comet when compared with the values found
by Cochran {\it et al.} (1992).  Since other groups use slightly
different scale lengths and fluorescence efficiencies, it is not
easy to compare with them.  However, there has never
been any evidence that Tempel 1 is not normal.

As noted above, the gas production rates of the various
species decreased from 1983 to 2005 and different species decreased
by various amounts.  Despite that, Fig.~\ref{ratios} shows no evidence
for a change in the ratios of species.  
However, only OH appeared to decrease from 1983 to 2005 by a factor much 
different than that for CN so it is not surprising that the ratios did not
change; for OH, there are not many points other than in 2005.
There are far more data points
from 2005 than from either 1983 or 1994 so the 2005 data dominate
the average values shown in each panel.  To check for subtle
changes in the ratios, we averaged each production rate ratio
by year.  These are listed in Table~\ref{avgratios}.  Inspection
of this Table shows that there are no real relative compositional differences
seen, within the errors.
\begin{table}
\caption{Production Rate Ratios by Year}\label{avgratios}
\vspace*{10pt}
\centering
\begin{tabular}{crrrrrr}
\hline
Year &
\multicolumn{1}{c}{$log \frac{Q(OH)}{Q(CN)}$} &
\multicolumn{1}{c}{$log \frac{Q(NH)}{Q(CN)}$} &
\multicolumn{1}{c}{$log \frac{Q(C_{3})}{Q(CN)}$} &
\multicolumn{1}{c}{$log \frac{Q(CH)}{Q(CN)}$} &
\multicolumn{1}{c}{$log \frac{Q(C_{2})}{Q(CN)}$} &
\multicolumn{1}{c}{$log \frac{Q(C_{2})}{Q(CN)}$} \\
 & & & & &
\multicolumn{1}{c}{$\Delta v=1$} & \multicolumn{1}{c}{$\Delta v=0$} \\
\hline
1983 &                &                & $-0.77\pm0.14$ & $0.17\pm0.10$ & $0.01\pm0.10$ & $0.11\pm0.14$ \\
1994 &  $2.29\pm0.18$ & $0.48\pm0.03$  & $-0.61\pm0.07$ & $0.44\pm0.26$ & $-0.03\pm0.06$ & $0.14\pm0.04$ \\
2005 &  $2.19\pm0.12$ & $0.47\pm0.11$  & $-0.58\pm0.18$ & $0.30\pm0.20$ & $0.05\pm0.07$ & $0.07\pm0.10$ \\
All &   $2.20\pm0.12$ & $0.48\pm0.11$  & $-0.61\pm0.18$ & $0.30\pm0.20$ & $0.04\pm0.07$ & $0.08\pm0.11$ \\
\hline
\end{tabular}

\end{table}

The data from the impact and post-impact are the circles at
1.5\,{\sc au}.  The clear circle is the night of impact.
We see that these production rate ratios from impact and after do not look
any different than the data from any other earlier night.  Thus,
we can conclude that the material populating the coma and observed in our
post-impact spectra has the same relative composition as the
normal coma material.

\section{Discussion}
The goal of the Deep Impact spacecraft mission was to study the interior
of a comet by excavating material from deep within the nucleus.
Groussin {\it et al.} (2007) analyzed IR spectra obtained from the flyby 
spacecraft to determine that the thermal inertia was low, ``most probably
$<50$ W K$^{-1}$ m$^{-2}$ s$^{1/2}$". This low thermal inertia implies that the
sublimation of volatiles generally occurs close to the surface.  The exact
depth of the crater produced by the impact is unknown, but probably included
``deep" materials (Schultz {\it et al.} 2007).
\nocite{groussinetal2007, schultzetal2007}

This unique event was monitored by observatories world-wide as a
complement to the data which were obtained by the spacecraft (Meech {\it et al.}
2005). \nocite{medeepimpact}
In order to understand the event, it was necessary to place the
impact monitoring observations into the context of the usual behavior
of this comet.  Our observations, reported in this paper, allow not
only for an understanding of the state of the comet at the time of
the impact, but also allow for an understanding of the natural
changes we see in this comet.

Comet Tempel 1 rotates relatively slowly, with a rotation period
of 1.701$\pm$0.014 days (A'Hearn {\it et al.} 2005). \nocite{ahdeepimpact}
Thus, our observations on a single night and a single slit orientation
were never more than 5\% of the rotation period.  Thus, even if there
had been a strong active region, it would not change position much relative
to our slit during a single night.  Major jets show up as moving ``bumps" in
the gas distribution as the material flows outward (Cochran and Trout 1994).
\nocite{cotr94}
Observations
on subsequent nights might have viewed different regions, 
but we do not see any differences from night-to-night
in our data.  In addition to the smoothly varying rotational modulation,
the comet would undergo sporadic outbursts (A'Hearn {\it et al.} 2005;
Meech {\it et al.} 2005; Farnham {\it et al.} 2007).
\nocite{fadeepimpact}  However, according to Table~1 of Farnham {\it et al.},
there were no outbursts on our dates of observations.  Thus, the
outbursts would have little affect on our observations.

Therefore, we can use our observations to comment on the properties
of the gas production of comet Tempel 1 during the season leading
up to the impact and also over the preceding two decades.
Overall, comet Tempel 1 is a moderate producer of gas.  
IR observations with the Deep Impact
spacecraft found that water ice ``is restricted to three discrete
and relatively small areas" (Sunshine {\it et al.} 2007) 
(this is not altogether consistent with the conclusion of Ferrin (2007)
that Tempel 1 is a young comet).
\nocite{sunshine2007,fe2007}
Our values for the production rates of the various optically observed
emissions are in good agreement with Schleicher (2007), with
small differences due to differences in fluorescence efficiencies,
scale lengths and outflow velocities used in the modeling.
Lara {\it et al.} (2006) also observed comet Tempel 1 during 2005.
Their derived production rates are lower than ours and Schleicher's
by a factor of a few for C$_{2}$, C$_{3}$, and CN, the species
they observed.  We cannot explain this difference.
\nocite{laetal06}

\begin{figure}
\vspace{4in}
\includegraphics{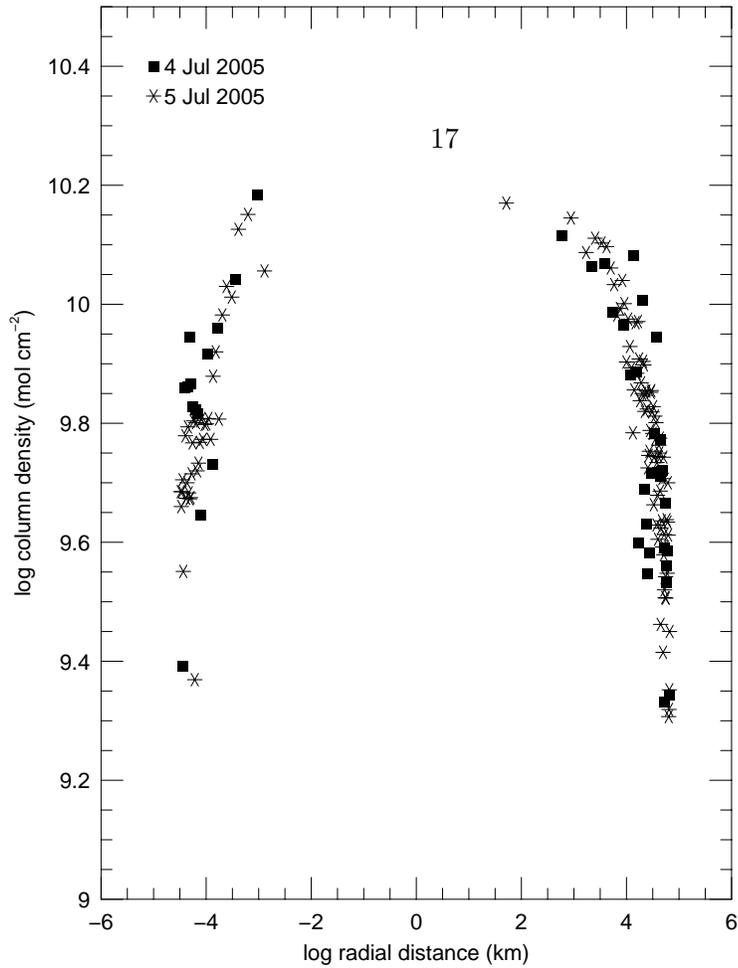}
\caption[fig3]{The column densities from 4 and 5 July 2005 UT
are compared, with the data from 4 July scaled upwards to match the level
of the data from 5 July.  Note that the distribution of the gas with
cometocentric distance is not the same on the east (negative direction) and
west sides of the comet.  The slit position angle was 20 degrees off of the
extended heliocentric radius vector so the slit is essentially along
the Sun-comet line.  Within the scatter of the 4 July data, there are no
difference in the profiles for 4 and 5 July on either side of the optocenter.
}\label{distrib}
\end{figure}

Within the error bars, the comet produced the same {\it relative} amounts
of all of the species that are in our bandpass for the three well
observed apparitions, even while the overall
amount of gas was decreasing. 
We see the same abundance ratios
in the spectra from 4 July 2005, the night of impact.  
The impact was an inherently non-steady state event while the Haser model
assumes steady state production.  Therefore, the production rates for July
2005 must be used cautiously.  On 4 July, the observations were obtained with
the first 40 minutes after the impact so ejecta from the impact would only
affect our inner few pixels (at 0.5 km sec$^{-1}$ ouitflow, gas from the crater will flow outwards 1200km in 40 minutes, or less than 1 pixel).  The outer part of the slit would still have the
ambient signal and this is what we would be measuring predominantly.  By 24
hours later, the ejected material would have traveled across our slit.  The
increase in production resulted from freshly exposed surface (Jackson
{\it et al.} 2008, in preparation).  Inspection of
the gas distribution shows that within our slit, the gas distribution has the
same general shape on 4 through 6 July as on other nights (Fig~\ref{distrib}). 
Rauer {\it et al.} (2006) found a ''bump" of material in observations from
4 and 5 July.  However, along the Sun-comet line (their Fig. 4), their profiles
look the same for the two nights. Within the scatter, ours do too, as seen
in Fig. 4.  We did not have a similarly oriented slit angle on 6 July.
\nocite{raueretal2006}
Thus, while a
Haser model production rate does not account for the impulsive nature of the
event, it allows for comparison with previous data.
The freshly released material has the same relative composition as the ambient
material.
As long as the impact
excavated material deep enough that it reached fresh, unprocessed
material, the implication of our findings is that the comet 
does not differentially lose one kind of ice versus another
as it passes through the inner Solar System repeatedly.  This also
means that prior studies of cometary composition via observations
of the normal coma emissions are probing the original composition
of these primitive bodies.

This rich data set, which spans more than two decades and four
apparitions allows us to look at secular changes in the comet.
We find that the total gas abundances declined from 1983 to 2005,
with decreases of factors of 1.3--2.7, depending on the species.
This is in good agreement with Schleicher (2007).  Except in
cases where a comet was totally disrupted (e.g. 1999 S4 (LINEAR)) or
its orbit changed significantly (e.g. 81P/Wild 2), such large
changes of gas production have not been seen before.  The simplest
explanation is that some region which was active on the surface of
the comet in 1983 was not well illuminated in 2005.  However,
A'Hearn {\it et al.} (2005) found that the obliquity of the comet
was only 11$^\circ$ and the viewing geometry was very similar in
1983, 1994 and 2005.  Thus, the only way the illumination of
a particular active region could change would be if that
active region was very near a pole.  This cannot be confirmed
with the flyby data since the encounter period was very short.  Such a
polar jet has been seen in comet Borrelly (Soderblom {\it et al.} 2002).
\nocite{soderetal:ds1}  Other mechanisms, such as the ``sealing" of
an active region from the fall-back of dust after perihelion,
cannot be ruled out.

We found that the gas production was not symmetric around
perihelion but peaked almost two months prior to perihelion.
Because we observed only at monthly intervals, we cannot
comment on whether different species peaked at different
times, something that was noted by Schleicher (2007).  
Such asymmetries are common in cometary activity and generally
indicate that small regions of the surface are active and
see changing illumination in the course of their orbit.
Indeed, as noted above, the Deep Impact spacecraft observations
did not observe distinct icy regions as sources of the activity.

Overall, we find that comet Tempel 1 has abundance patterns which
are similar to the vast majority of comets and can be classified
as a ``normal" comet.  These abundance patterns do not change
with depth, as evidenced from the post-impact spectra. 
Therefore, as this comet continues to evolve, it is unlikely
to change its abundances patterns.

\section{Summary}
Our main findings are that: 
\begin{itemize}
\item The relative abundances of
different gas species are the same, within the errors, for all three
apparitions.
\item The comet showed a generalized decrease
in activity from the 1983 apparition to the 2005 apparition.
The amount of decrease varied, depending on the species, from a factor
of 1.3 to a factor of 2.7. 
\item The production of gas peaks more than a month prior to perihelion. 
\item Comet 9P/Tempel 1 is like the great majority of comets
with ``normal" gas abundance ratios.
\end{itemize}

Deep Impact was a unique experiment to explore the interior of
a comet by blasting free material and analyzing this newly
released material.  In this paper, we have described two decades
of observations using spectrographs at McDonald Observatory.
These observations allow for placing the Deep Impact results
into context with the normal behavior of this comet.

\vspace*{0.75in}
\begin{center}Acknowledgments\end{center}
This research was supported by NASA Grant NNG04G162G and predecessor
grants.  McDonald Observatory is operated by The University of
Texas at Austin.

\newpage

\end{document}